\input harvmac

 \def\frac#1#2{{#1\over #2}}
 
 \def\s{\sqrt}
 \def\p{\partial}

 \def\al{\alpha'}
 \def\de{\partial}

 \def\f {\frac}
 \def\ti{\tilde}
 \def\ap{\alpha}

 \def\ddd{\cdot\cdot\cdot}

 \def\lb{\rangle}
 \def\ep{\epsilon}

 \def\vp{\varphi}
 \def\tvp{\tilde{\varphi}}

\def\bp{\bar{\partial}}

\font\cmss=cmss10
\font\cmsss=cmss10 at 7pt

\def\inbar{\vrule height1.5ex width.4pt depth0pt}
\def\IC{{\relax\,\hbox{$\inbar\kern-.3em{\rm C}$}}}
\def\IR{\relax{\rm I\kern-.18em R}}
\def\IZ{\relax\ifmmode\mathchoice
{\hbox{\cmss Z\kern-.4em Z}}{\hbox{\cmss Z\kern-.4em Z}}
{\lower.9pt\hbox{\cmsss Z\kern-.4em Z}}
{\lower1.2pt\hbox{\cmsss Z\kern-.4em Z}}\else{\cmss Z\kern-.4em
Z}\fi}

\lref\VafaWX{
  C.~Vafa,
  ``Modular Invariance And Discrete Torsion On Orbifolds,''
  Nucl.\ Phys.\ B {\bf 273}, 592 (1986).
}

\lref\MelvinQX{
  M.~A.~Melvin,
  ``Pure Magnetic And Electric Geons,''
  Phys.\ Lett.\  {\bf 8}, 65 (1964).
}

\lref\RussoCV{
  J.~G.~Russo and A.~A.~Tseytlin,
  ``Constant magnetic field in closed string theory: an exactly solvable
  model,''
  Nucl.\ Phys.\ B {\bf 448}, 293 (1995)
  [arXiv:hep-th/9411099].
}

\lref\RussoTJ{
  J.~G.~Russo and A.~A.~Tseytlin,
  ``Exactly solvable string models of curved space-time backgrounds,''
  Nucl.\ Phys.\ B {\bf 449}, 91 (1995)
  [arXiv:hep-th/9502038].
}

\lref\LeighEP{
  R.~G.~Leigh and M.~J.~Strassler,
  ``Exactly marginal operators and duality in four-dimensional N=1
  supersymmetric gauge theory,''
  Nucl.\ Phys.\ B {\bf 447}, 95 (1995)
  [arXiv:hep-th/9503121].
}

\lref\RussoIK{
  J.~G.~Russo and A.~A.~Tseytlin,
  ``Magnetic flux tube models in superstring theory,''
  Nucl.\ Phys.\ B {\bf 461}, 131 (1996)
  [arXiv:hep-th/9508068].
}

\lref\DouglasSW{
  M.~R.~Douglas and G.~W.~Moore,
  ``D-branes, Quivers, and ALE Instantons,''
  arXiv:hep-th/9603167.
}

\lref\DiaconescuBR{
  D.~E.~Diaconescu, M.~R.~Douglas and J.~Gomis,
  ``Fractional branes and wrapped branes,''
  JHEP {\bf 9802}, 013 (1998)
  [arXiv:hep-th/9712230].
}

\lref\DouglasXA{
  M.~R.~Douglas,
  ``D-branes and discrete torsion,''
  arXiv:hep-th/9807235.
}

\lref\DouglasHQ{
  M.~R.~Douglas and B.~Fiol,
  ``D-branes and discrete torsion II,''
  arXiv:hep-th/9903031.
}

\lref\DiaconescuDT{
  D.~E.~Diaconescu and J.~Gomis,
  ``Fractional branes and boundary states in orbifold theories,''
  JHEP {\bf 0010}, 001 (2000)
  [arXiv:hep-th/9906242].
}

\lref\SeibergVS{
  N.~Seiberg and E.~Witten,
  ``String theory and noncommutative geometry,''
  JHEP {\bf 9909}, 032 (1999)
  [arXiv:hep-th/9908142].
}

\lref\BerensteinHY{
  D.~Berenstein and R.~G.~Leigh,
  ``Discrete torsion, AdS/CFT and duality,''
  JHEP {\bf 0001}, 038 (2000)
  [arXiv:hep-th/0001055].
}

\lref\GomisEJ{
  J.~Gomis,
  ``D-branes on orbifolds with discrete torsion and topological obstruction,''
  JHEP {\bf 0005}, 006 (2000)
  [arXiv:hep-th/0001200].
}

\lref\BerensteinUX{
  D.~Berenstein, V.~Jejjala and R.~G.~Leigh,
  ``Marginal and relevant deformations of N = 4 field theories and
  non-commutative moduli spaces of vacua,''
  Nucl.\ Phys.\ B {\bf 589}, 196 (2000)
  [arXiv:hep-th/0005087].
}

\lref\GaberdielFE{
  M.~R.~Gaberdiel,
  ``Discrete torsion orbifolds and D-branes,''
  JHEP {\bf 0011}, 026 (2000)
  [arXiv:hep-th/0008230].
}

\lref\CostaNW{
  M.~S.~Costa and M.~Gutperle,
  ``The Ka\l u\.{z}a-Klein Melvin solution in M-theory,''
  JHEP {\bf 0103}, 027 (2001)
  [arXiv:hep-th/0012072].
}

\lref\CrapsXW{
  B.~Craps and M.~R.~Gaberdiel,
  ``Discrete torsion orbifolds and D-branes II,''
  JHEP {\bf 0104}, 013 (2001)
  [arXiv:hep-th/0101143].
}

\lref\GutperleMB{
  M.~Gutperle and A.~Strominger,
  ``Fluxbranes in string theory,''
  JHEP {\bf 0106}, 035 (2001)
  [arXiv:hep-th/0104136].
}

\lref\AdamsSV{
  A.~Adams, J.~Polchinski and E.~Silverstein,
  ``Don't panic! Closed string tachyons in ALE space-times,''
  JHEP {\bf 0110}, 029 (2001)
  [arXiv:hep-th/0108075].
}

\lref\TakayanagiJJ{
  T.~Takayanagi and T.~Uesugi,
  ``Orbifolds as Melvin geometry,''
  JHEP {\bf 0112}, 004 (2001)
  [arXiv:hep-th/0110099].
}

\lref\RussoNA{
  J.~G.~Russo and A.~A.~Tseytlin,
  ``Supersymmetric fluxbrane intersections and closed string tachyons,''
  JHEP {\bf 0111}, 065 (2001)
  [arXiv:hep-th/0110107].
}

\lref\DudasUX{
  E.~Dudas and J.~Mourad,
  ``D-branes in string theory Melvin backgrounds,''
  Nucl.\ Phys.\ B {\bf 622}, 46 (2002)
  [arXiv:hep-th/0110186].
}

\lref\TakayanagiAJ{
  T.~Takayanagi and T.~Uesugi,
  ``D-branes in Melvin background,''
  JHEP {\bf 0111}, 036 (2001)
  [arXiv:hep-th/0110200].
}

\lref\DavidVM{
  J.~R.~David, M.~Gutperle, M.~Headrick and S.~Minwalla,
  ``Closed string tachyon condensation on twisted circles,''
  JHEP {\bf 0202}, 041 (2002)
  [arXiv:hep-th/0111212].
}

\lref\TakayanagiGU{
  T.~Takayanagi and T.~Uesugi,
  ``Flux stabilization of D-branes in NSNS Melvin background,''
  Phys.\ Lett.\ B {\bf 528}, 156 (2002)
  [arXiv:hep-th/0112199].
}

\lref\GregoryYB{
  R.~Gregory and J.~A.~Harvey,
  ``Spacetime decay of cones at strong coupling,''
  Class.\ Quant.\ Grav.\  {\bf 20}, L231 (2003)
  [arXiv:hep-th/0306146].
}

\lref\CornalbaKD{
  L.~Cornalba and M.~S.~Costa,
  ``Time-dependent orbifolds and string cosmology,''
  Fortsch.\ Phys.\  {\bf 52}, 145 (2004)
  [arXiv:hep-th/0310099].
}

\lref\HeadrickYU{
  M.~Headrick,
  ``Decay of $C/Z_n$: Exact supergravity solutions,''
  JHEP {\bf 0403}, 025 (2004)
  [arXiv:hep-th/0312213].
}

\lref\HashimotoVE{
  A.~Hashimoto and L.~Pando Zayas,
  ``Correspondence principle for black holes in plane waves,''
  JHEP {\bf 0403}, 014 (2004)
  [arXiv:hep-th/0401197].
}

\lref\HeadrickHZ{
  M.~Headrick, S.~Minwalla and T.~Takayanagi,
  ``Closed string tachyon condensation: An overview,''
  Class.\ Quant.\ Grav.\  {\bf 21}, S1539 (2004)
  [arXiv:hep-th/0405064].
}

\lref\LuninJY{
  O.~Lunin and J.~Maldacena,
  ``Deforming field theories with $U(1) \times U(1)$ global
  symmetry and their gravity duals,''
  JHEP {\bf 0505}, 033 (2005)
  [arXiv:hep-th/0502086].
}

\lref\FrolovDJ{
  S.~Frolov,
  ``Lax pair for strings in Lunin-Maldacena background,''
  JHEP {\bf 0505}, 069 (2005)
  [arXiv:hep-th/0503201].
}

\lref\RussoYU{
  J.~G.~Russo,
  ``String spectrum of curved string backgrounds obtained by T-duality and
  shifts of polar angles,''
  arXiv:hep-th/0508125.
}

\Title
{\vbox{
 \baselineskip12pt
\hbox{hep-th/0509036}
\hbox{NSF-KITP-05-72}
\hbox{MCTP-05-92}
}}
{\vbox{
\centerline{String Theory in $\beta$ Deformed Spacetimes}
}}

\centerline{
Marcus Spradlin\foot{On leave from the Department of Physics and
the Michigan Center for
Theoretical Physics, University of Michigan, Ann Arbor, MI 48109 USA.}${}^{,2}$,
Tadashi Takayanagi
and Anastasia Volovich\foot{Address after September 15:
School of Natural Sciences, Institute for Advanced Study, Princeton NJ
08540 USA.}
}

\bigskip
\centerline{Kavli Institute for Theoretical Physics}
\centerline{University of California}
\centerline{Santa Barbara, CA 93106 USA}
\centerline{\tt spradlin, takayana, nastja@kitp.ucsb.edu}

\vskip .3in \centerline{\bf Abstract}
Fluxbrane-like backgrounds obtained from flat space by a sequence
of T-dualities and shifts of polar coordinates ($\beta$ deformations)
provide an
interesting class of exactly solvable string theories. We compute
the one-loop partition function for various such deformed spaces
and study their spectrum of D-branes. For rational values of the
$B$-field these models are equivalent to $\IZ_N \times \IZ_N$ orbifolds
with discrete torsion.
We also obtain an interesting new class of time-dependent
backgrounds which resemble localized closed string tachyon condensation.

\Date{September 2005}

\listtoc
\writetoc

\newsec{Introduction and Summary}

String theory often requires modified notions of geometry. This is
not only because the string world-sheet has a finite length scale,
but also because a background for string theory is specified by 
NSNS and RR fluxes
in addition to the metric.
In this paper we study an interesting example of a closed string
background with NSNS flux \LuninJY,
\eqn\backg{\eqalign{
ds^2&=dr_1^2+dr_2^2+\f{r_1^2}{1+b^2r_1^2r_2^2}d\vp_1^2+
\f{r_2^2}{1+b^2r_1^2r_2^2}d\vp_2^2,  \cr
e^{2(\phi-\phi_0)}&=\f{1}{1+b^2r_1^2r_2^2}, ~~~
B_{\vp_1\vp_2}=-\f{b r_1^2r_2^2}{1+b^2r_1^2r_2^2}}} and
some generalizations thereof. This background shares some similarities
with the Melvin fluxbrane solution~\refs{\MelvinQX,\CostaNW,\GutperleMB}.
In particular, like
the Melvin case \refs{\RussoCV,\RussoTJ,\RussoIK}, string theory in~\backg\
is exactly solvable~\RussoYU. In this case the solvability is due
to the fact that~\backg\ is actually equivalent to string theory
on $\IR^4$ after a T-duality, linear redefinition of the polar
coordinates, followed by another T-duality. A sequence of such
TsT dualities was recently used in~\LuninJY\ to find the
supergravity solution dual to a certain marginal deformation of
${\cal N} = 4$ Yang-Mills theory \refs{\LeighEP,\BerensteinUX}
called the $\beta$ deformation,
but the procedure of using sequences of TsT transformations can of
course be used more generally to find interesting new curved
backgrounds which are solutions of the string equations of motion to all
orders in $\alpha'$.

When both radii $r_i$ are large  $(br_1r_2\gg 1)$, the
metric~\backg\ approaches \eqn\asymptotic{ ds^2 \approx dr_1^2 +
dr_2^2 + {1 \over b^2 r_1^2} d \varphi_2^2 + {1 \over b^2 r_2^2} d
\varphi_1^2 = dr_1^2 + dr_2^2 + { \beta^2 r_1^2} d
{\varphi}_2^{\prime 2} + { \beta^2 r_2^2} d {\varphi}_1^{\prime 2}, }
where we have performed T-duality $\varphi_i \to {\varphi}^\prime_i$ on
the shrinking $\varphi_i$ circles and we have defined $\beta \equiv
\alpha' b$.
The form of the asymptotic metric~\asymptotic\
suggests that for rational $\beta$ this background should be
related to an orbifold of $\IC^2$, and we will see  below that this
is indeed the case. In particular, the bosonic string in~\backg\
at $\beta = k/N$ is equivalent to the orbifold $\IC^2/\IZ_N \times
\IZ_N$ with discrete torsion \VafaWX. From this equivalence we can
understand the effect of discrete torsion in a geometrical way by
examining the solution \backg. Intuitively, we can say that the
discrete torsion smooths out the conical singularity
otherwise  present in the
orbifold $\IC^2/\IZ_N \times \IZ_N$.

Superstrings on~\backg\ also give various orbifolds with discrete
torsion, though the classification is more refined.  We find that
superstring theory on~\backg\ is equivalent to
\eqn\results{\eqalign{ \IC^2/\IZ_N \times \IZ_N \qquad & {\rm for}
\qquad \beta = 4 k/N, \cr \IC^2/\IZ_{2N} \times \IZ_{2N} \qquad &
{\rm for} \qquad \beta = (4k+2)/N, \cr \IC^2/\IZ_{4N} \times
\IZ_{4N} \qquad & {\rm for} \qquad \beta = (4k+1)/N {\rm ~or~}
(4k+3)/N. }}
Superstring theory on such orbifolds has
localized tachyons pinned to the orbifold points~\AdamsSV, and
we will also see explicitly below that 
some states come from a type 0 GSO projection (i.e. there are
(NS$-$,NS$-$) sectors), so there are also bulk tachyons as well.
In fact, superstring theory on~\backg\ has closed string tachyons
for any value of $\beta$ (not only rational)\foot{For example,
in the notation of
\RussoYU, states with $\nu_2 = 0$ and only zero-mode excitations
have mass $\alpha' M^2 = - 2 \nu_1 (J_{1R}  - J_{1L})$, which can have
either sign. We are grateful to A.~Adams for pointing out the existence of
such tachyonic states in the orbifold limit.}
and might therefore provide an 
interesting laboratory for studying closed string tachyon
condensation, as the Melvin solution certainly has
\refs{\GutperleMB,\DavidVM}.

The modified geometry implied by string theory can also be studied by
using D-branes as probes.  
We study the spectrum of D-branes in the background~\backg, as has
been done in the Melvin case in
\refs{\TakayanagiAJ,\DudasUX}.
In the orbifold limit when $\beta$ is rational, the theory has both
bulk D-branes and fractional D-branes, and we show agreement with
results expected from the study of
D-branes on orbifolds with discrete torsion \refs{\DouglasXA \DouglasHQ 
\BerensteinHY
\GomisEJ \GaberdielFE-\CrapsXW}.

We also study a generalization
of~\backg\ involving a three parameter deformation of $\IC^3$ \refs{\LuninJY,
\RussoYU,\FrolovDJ}
which generically has tachyons but is supersymmetric when all
three parameters are equal.
Another generalization of~\backg\ gives an interesting new time-dependent
background which has some resemblence to the decay of localized closed
string tachyons \AdamsSV\ (see \HeadrickHZ\ for a review and references
therein).

The outline of this paper is as follows.  We review the TsT transformation
and the solution of the string sigma model for~\backg\ in section 2.
In section 3 we calculate, in both the operator and path-integral
formalisms, the one-loop partition functions for bosonic and superstring
theory in this background.  We show that for rational $\beta$ they
agree with the various orbifolds with discrete torsion listed above.
In section 4 we classify the various D-branes in~\backg\ and compute
boundary states for some of them, demonstrating agreement with expected
results in the orbifold limits.  In section 5 we summarize the conclusions
of similar analysis for the three parameter deformation of $\IC^3$, and
finally in section 6 we comment on the time-dependent version of this
background.

\newsec{Solution of the Closed String Theory}

In this section we review the solution of string theory in
the background~\backg, following~\RussoYU.
The sigma model describing closed strings in this background is
\eqn\sigmaone{\eqalign{S &= {1 \over
2 \pi \alpha'} \int d^2z\ \Bigl{[} \de r_1\bar{\de}r_1
+ \de r_2\bar{\de}r_2+\f{r_1^2}{1+b^2r_1^2r_2^2}\de \vp_1\bar{\de}\vp_1+
\f{r_2^2}{1+b^2 r_1^2r_2^2}\de \vp_2\bar{\de}\vp_2\cr
&\qquad\qquad\qquad
+\f{b r_1^2 r_2^2}{1+b^2 r_1^2r_2^2}(\de\vp_2\bar{\de}\vp_1-\de\vp_1\bar{\de}\vp_2)
\Bigr{]}.}}
We first T-dualize in $\vp_1 \to \tilde{\vp}_1$, finding
\eqn\sigmatwo{
S=
{1 \over 2 \pi \alpha'}
\int d^2 z\ \left[
\de r_1\bar{\de} r_1+ \de r_2\bar{\de} r_2+\f{1}{r^2_1}\de\tvp_1\bar{\de}\tvp_1
+r_2^2(\de\vp_2+b\de\tvp_1)(\bar{\de}\vp_2+b\bar{\de}\tvp_1)\right].
}
Then we define the new coordinate ${\phi}_2={\vp}_2+b\tvp_1$ and
T-dualize again in $\tvp_1 \to \phi_1$.
This gives
\eqn\sigmathree{S= {1 \over 2 \pi \alpha'} \int d^2z
\left[ \de r_1\bar{\de}r_1 + \de r_2\bar{\de}r_2+r_1^2\de
\phi_1\bar{\de}\phi_1+r_2^2\de \phi_2\bar{\de}\phi_2\right].
}
Recognizing~\sigmathree\ as the sigma model for $\IC^2$ written
in polar coordinates, we define the complex free fields
$X_1=r_1e^{i\phi_1}$ and
$X_2=r_2e^{i\phi_2}$, whose action is just
\eqn\actions{S= {1 \over 2 \pi \alpha'} \int
d^2z \left[ \de X_1\bar{\de}\bar{X}_1+\de
X_2\bar{\de}\bar{X}_2 \right].
}

When two sigma models are related to each other by a T-duality in
$X \to \tilde{X}$, their classical solutions are related by the
formulas (see appendix A)
\eqn\tduality{\eqalign{ \p \tilde{X}_\mu &=   B_{\mu\nu}(X) \p
X^\nu - G_{\mu \nu}(X) \p X^\nu,\cr \bp \tilde{X}_\mu &=  B_{\mu\nu}(X)
\bp X^\nu + G_{\mu \nu}(X) \bp X^\nu. }} If we define
$\tilde{\varphi}_2$ to satisfy $\phi_1 = \varphi_1 - b
\tilde{\varphi}_2$, in analogy with the definition $\phi_2 =
\varphi_2 + b \tilde{\varphi}_1$ we used above, then we find that
the equations~\tduality\ imply that the relations
\eqn\relaf{\eqalign{\de\tvp_{i}&=-r_{i}^2\de\phi_{i}=\f{i}{2}[\bar{X}_{i}\de
X_{i}-X_{i}\de \bar{X}_{i}], \cr
\bar{\de}\tvp_{i}&=r_{i}^2\bar{\de}\phi_{i}=-\f{i}{2}[\bar{X}_{i}\bar{\de}
X_{i} -X_{i}\bar{\de} \bar{X}_{i}] }} hold on-shell for $i=1,2$.
Integrating the relation~\relaf\ over $\sigma$ leads to the
boundary condition
\eqn\perone{\tvp_{i}(t,\sigma+\pi)=\tvp_{i}(t,\sigma)-2\pi\al
(J_{i}+\tilde{J}_{i}), \qquad i= 1,2,} where we have introduced
the familiar left- and right-moving angular momentum operators
\eqn\jch{\eqalign{ J_{i}&=\f{i}{4\pi\al}\int^{\pi}_0d\sigma
[X_{i}\de\bar{X}_{i}-\bar{X}_{i}\de X_{i}], \cr \tilde{J}_i &=
\f{i}{4\pi\al}\int^{\pi}_0d\sigma
[X_{i}\bp\bar{X}_{i}-\bar{X}_{i}\bp X_{i}], \qquad i = 1,2. }}
{}From~\relaf\ it follows that closed string boundary conditions
\eqn\aaa{
\varphi_i(t, \sigma + \pi) = \varphi_i(t, \sigma), \qquad i=1,2
}
for the original sigma model~\sigmaone\ translate into
\eqn\aaa{\eqalign{
\phi_1(t,\sigma+\pi) &= \phi_1(t,\sigma) + 2 \pi \beta (J_2
+ \tilde{J}_2), \cr
\phi_2(t,\sigma+\pi) &= \phi_2(t,\sigma) - 2 \pi \beta (J_1
+ \tilde{J}_1).
}}

We conclude from this analysis that the sigma model~\sigmaone\
reduces to two free complex bosons with the twisted boundary
conditions \eqn\pertw{\eqalign{X_1(t,\sigma+\pi) &=e^{2\pi i \beta
(J_{2}+\tilde{J}_{2})}X_1(t,\sigma), \cr X_2(t,\sigma+\pi)
&=e^{-2\pi i \beta (J_{1}+\tilde{J}_{1})}X_2(t,\sigma).}} Note
that since the angular momenta $J_i$ are integer quantized, it is
manifest from~\pertw\ that the theory is periodic under $\beta \to
\beta + 1$ in the bosonic string case. We can also consider the
supersymmetric completion of~\sigmaone, in which case the
periodicity becomes $\beta \sim \beta + 2$, as we will see
explicitly below.

\newsec{One-Loop Partition Functions}

The one-loop vacuum amplitude in closed string theory can be
expressed as \eqn\oneloop{ Z_{T^2}=iV_{D-4} \int \f{d\tau
d\bar{\tau}}{4\tau_2} {1 \over (4\pi^2 \alpha' \tau_2)^{\ha(D-4)}}
Z(\tau,\bar{\tau},\beta), } where $V_{D-4}$ is the volume of the
non-compact dimensions described by a trivial CFT which we omit in
the discussion below, and $Z(\tau,\bar{\tau},\beta)$ is the 4D
partition function which we are interested in. We take $D=10$ for the
superstring and $D=26$ for the bosonic string.

\subsec{Operator Formalism}

The partition function $Z$
can be computed in the operator formalism as
\eqn\Zdef{
Z(\tau,\bar{\tau},\beta) = \Tr[q^{L_0} \bar{q}^{\tilde{L}_0}],
}
with $q = e^{2 \pi i \tau}$ and $\tau = \tau_1 + i \tau_2$.
In~\RussoYU\ it was shown that the canonical quantization of the
model~\sigmaone\ leads to
the expressions
\eqn\Lzerodef{\eqalign{
L_0 &= N - a
- (\nu_1 - [\nu_1]) J_1 - (\nu_2 - [\nu_2]) J_2,
\cr
\tilde{L}{}_0 &= \tilde{N} - \tilde{a}
+ (\nu_1 - [\nu_1]) \tilde{J}_1 + (\nu_2 - [\nu_2]) \tilde{J}_2,
}}
where $a$ and $\tilde{a}$ are the usual normal-ordering constants,
$[x]$ denotes the greatest integer less than or equal to $x$, and
\eqn\nudefs{
\nu_1 = \beta (J_2 + \tilde{J}_2), \qquad
\nu_2 = - \beta (J_1 + \tilde{J}_1).
}

In order to calculate the trace~\Zdef\ we find
it convenient to follow a similar calculation
performed in~\refs{\RussoCV,\RussoTJ,\HashimotoVE}.
We begin by inserting delta-functions to write
\eqn\stepone{
Z = \int d^2 j_1 d^2 j_2\
\Tr[q^{L_0} \bar{q}^{\tilde{L}_0}
\delta^2(J_1 - j_1) \delta^2(J_2 - j_2)].
}
These enable us to set $J_i = j_i$ and $\tilde{J}_i = \bar{\jmath}_i$ inside
the trace.  With the help of the formula
\eqn\deltafun{
\delta^2(z) = \int d^2 \chi\ e^{2 \pi i \chi z + 2 \pi i \bar{\chi} \bar{z}}
}
we then obtain
an expression in which the left- and right-moving traces factorize,
\eqn\steptwo{\eqalign{
Z &= \int d^2 j_1d^2 j_2 d^2 \chi_1 d^2 \chi_2\
q^{ - (\nu_1 - [\nu_1]) j_1 - (\nu_2 - [\nu_2]) j_2}
\bar{q}^{(\nu_1 - [\nu_1]) \bar{\jmath}_1
+ (\nu_2 - [\nu_2]) \bar{\jmath}_2}
\cr
&\times
e^{-2 \pi i (\chi_1 j_1 + \bar{\chi}_1 \bar{\jmath}_1
+ \chi_2 j_2 +  \bar{\chi}_2 \bar{\jmath}_2)}
\Tr[q^{N - a} e^{2 \pi i \chi_1 J_1 + 2 \pi i \chi_2 J_2}]
\Tr[\bar{q}^{\tilde{N} - \tilde{a}} e^{2 \pi i \bar{\chi}_1 \tilde{J}_1
+ 2 \pi i \bar{\chi}_2 \tilde{J}_2}].
}}

The result~\steptwo\ is equally valid for the bosonic string and
the superstring.  Let us consider first the superstring, in which
case the oscillator traces lead to (see appendix B)
\eqn\tracesone{ F(\chi_1,\chi_2,\tau) \equiv \left| {
\vartheta_1(\ha (\chi_1 + \chi_2)|\tau)^2 \vartheta_1(\ha(\chi_1 -
\chi_2)|\tau)^2 \over \eta(\tau)^6 \vartheta_1(\chi_1|\tau)
\vartheta_1(\chi_2|\tau)}\right|^2. } This result is correct for
the standard type II GSO projection.
 As shown in~\RussoYU, when
$[\nu_1] + [\nu_2]$ is odd one must use the opposite GSO
projection; we will incorporate this fact momentarily.
Even though
the partition function $\tracesone$ is expressed in the form which
emerges naturally from
the Green-Schwarz formalism, it is equivalent to the more familiar
expression in the NSR formalism via Jacobi's identity (see appendix B).

In~\steptwo\ we can perform the change of variables
\eqn\aaa{
\chi_i \to \chi_i + \tau [\nu_i], \qquad
\bar{\chi}_i \to \bar{\chi}_i + \bar{\tau} [\nu_i]
}
to eliminate the appearance of $[\nu]_i$.
The factor~\tracesone\ is invariant under this shift if
$[\nu_1] + [\nu_2]$ is even, but for $[\nu_1] + [\nu_2]$ odd we have
\eqn\tracestwo{
F(\chi_1 + [\nu_1] \tau, \chi_2 + [\nu_2] \tau, \tau)=
\left| {\vartheta_4(\ha(\chi_1+\chi_2)|\tau)^2
\vartheta_4(\ha(\chi_1-\chi_2)|\tau)^2 \over
\eta(\tau)^6 \vartheta_1(\chi_1|\tau) \vartheta_1(\chi_2|\tau)}\right|^2,
}
which is precisely the required
oscillator trace for the opposite GSO projection.
Finally we can perform the $j_i$ integrals to arrive at
\eqn\superstring{
Z = \int {d^2 \chi_1 d^2\chi_2\over (2 \beta \tau_2)^2}
e^{{\pi \over \tau_2 \beta} (\chi_1 \bar{\chi}_2 - \chi_2 \bar{\chi}_1)}
\left| {\vartheta_1(\ha(\chi_1+\chi_2)|\tau)^2
\vartheta_1(\ha(\chi_1-\chi_2)|\tau)^2 \over \eta(\tau)^6
\vartheta_1(\chi_1|\tau) \vartheta_1(\chi_2|\tau)}\right|^2,
}
which naturally incorporates the proper GSO projections.
This result for the one-loop amplitude can also be obtained by
performing the path integral, as we will show in the following subsection.
Plugging~\superstring\ into~\oneloop\ gives a modular
invariant amplitude in $D=10$.

The result~\superstring\ is actually periodic in $\beta$,
\eqn\betashift{
Z(\tau,\bar{\tau},\beta + 2) = Z(\tau,\bar{\tau},\beta).
}
In order to expose this symmetry, we use the fact that the
combination of $\vartheta$-functions appearing in~\superstring\ is invariant
under
\eqn\periodicity{
\chi_1 \to \chi_1 + 2 (m' - m \tau), \qquad
\chi_2 \to \chi_2 + 2 (n' - n \tau),
}
for $m,m',n,n' \in \IZ$.
We can exploit this periodicity to divide the
integral over $\chi_2$ into a sum over $n$ and $n'$ as well as an integral
over the torus $(0,0) \sim (2,2\tau)$,
\eqn\torusone{
Z = \sum_{n,n'\in \IZ}
\int{d^2 \chi_1 \over 2 \beta \tau_2}
\int_{T^2}
{d^2 \chi_2 \over 2 \beta \tau_2}
e^{  {\pi \over \tau_2 \beta}(\chi_1 \bar{\chi}_2 -
\chi_2 \bar{\chi_1})}
e^{{2 \pi \over \tau_2 \beta}[  \chi_1 (n' - n \bar{\tau}) -
 \bar{\chi}_1 (n' - n \tau)]}
F(\chi_1,\chi_2,\tau).
}
The sum over $n$ and $n'$ gives  the delta-functions
\eqn\aaa{
{\tau_2 \beta^2 \over 2}
\sum_{p,p' \in \IZ}
\delta^2( \chi_1 - \ha \beta (p' - p \tau))
}
which allow easy evaluation of the $\chi_1$ integral,
\eqn\resone{
Z = {1 \over 8 \tau_2}
\sum_{p,p' \in \IZ} \int_{T^2} {d^2 \chi_2}\
e^{{\pi \over 2 \tau_2}[(p'-p \tau) \bar{\chi}_2 -
(p'-p \bar{\tau})\chi_2]}
F(\ha \beta (p' - p \tau), \chi_2, \tau).
}
In this form the periodicity under $\beta \to \beta + 4$ is manifest,
and the finer periodicity~\betashift\ can be seen by shifting the $\chi_2$
integration variable by $\chi_2 \to \chi_2 + (p' - p \tau)$.

\subsec{Path-integral Formalism}

The torus amplitude~\superstring\ may also be obtained by
directly evaluating
 the one-loop functional determinant of the sigma model
action~\sigmaone. This calculation is facilitated by the
introduction of a useful choice of auxiliary fields, as in a
similar calculation for the Melvin background~\refs{\RussoCV,\RussoTJ}.
In our case we introduce two complex auxiliary fields $U$ and $V$
with the action
\eqn\pathtwo{\eqalign{S&={1 \over 2 \pi \alpha'} \int d^2z\
\Bigl[\de r_1\bar{\de} r_1+\de r_2\bar{\de} r_2
+r_2^2\de\vp_2\bar{\de}\vp_2+b^2r_2^2V\bar{V}
+ r_1^2 U \bar{U}
\cr
&\qquad\qquad\qquad\qquad
+\bar{V}(\de\vp_1+br_2^2\de\vp_2)-V(\bar{\de}\vp_1-br_2^2\bar{\de}\vp_2)
-U\bar{V}+\bar{U}V\Bigr].}}
It is straightforward to check that integrating out $U$ and $V$
recovers~\sigmaone.

If we instead start from~\pathtwo\ and perform
the path integral over $\varphi_1$, we find
the flatness condition $\p \bar{V} - \bp V = 0$
which is solved by writing
\eqn\Vdef{
V = v + \p \varphi, \qquad \bar{V} = \bar{v} + \bp \varphi
}
in terms of a complex constant (zero-mode) $v$ and a new field $\varphi$.
Actually, the flatness condition arises from integrating out only the
non-zero modes of $\varphi_1$.  There is a term in the action,
\eqn\totalde{
\int ( \bar{v} \p \varphi_1 - v \bp \varphi_1)
}
which
depends also on the zero-mode of $\varphi_1$, which is given by
\eqn\zeromode{
\varphi_1 = m \sigma_1 + {m' - m \tau_1 \over \tau_2} \sigma_2.
}
We will incorporate this zero-mode shortly.

The path integral over $V$ therefore reduces to an ordinary integral
over $v$ and a path integral over $\varphi$, although note that the latter has
no zero mode since only derivatives of $\varphi$ appear in~\Vdef.
We obtain
\eqn\paththree{\eqalign{S&={1 \over
2 \pi \alpha'}\int d^2z \Bigl[\de r_1\bar{\de}
r_1 +r_1^2 U\bar{U}-U(\bar{v}+\bar{\de}\vp)+\bar{U}(v+\de\vp)\cr &
\qquad\qquad\qquad\qquad
 +\de r_2\bar{\de}
r_2+r_2^2(bv+b\de\vp+\de\vp_2)(b\bar{v}+b\bar{\de}\vp+\bar{\de}\vp_2)
\Bigr].}} Next we change variables in the path integral from
$(\varphi,\varphi_2)$ to $(\varphi, \varphi_2 + b \varphi)$. Then
we can integrate out $\vp$ (which, recall, has no zero modes, so
we don't have to worry about total derivative terms such
as~\totalde) to obtain another flatness condition $\de
\bar{U}-\bar{\de}U=0$. Again we can write \eqn\ucond{U=u+\de
\theta,\ \ \ \bar{U}=\bar{u}+\bar{\de}\theta,} in terms of a
constant $u$ and a function $\theta$ with no zero-mode. Finally we
arrive at the  result \eqn\paththreee{\eqalign{S&= {1 \over 2 \pi
\alpha'} \int dz^2 \Bigl[\de r_1\bar{\de} r_1 +r_1^2(u+\de
\theta)(\bar{u}+\bar{\de}\theta)-u\bar{v}+\bar{u}v-\bar{v}\de\theta
+v\bar{\de}\theta\cr &\qquad\qquad\qquad\qquad +\de r_2\bar{\de}
r_2+r_2^2(bv+b\de\vp+\de\vp_2)(b\bar{v}+b\bar{\de}\vp+\bar{\de}\vp_2)
\Bigr].}} We can set the terms $\int(-\bar{v} \p \theta + v \bp
\theta)$ to zero because $\theta$ has no zero-modes, and then
rewrite this action in terms of the coordinates
$Z_1=r_1e^{i\theta}$ and $Z_2=r_2e^{i(\vp_2+b\vp)}$ as
\eqn\pathfin{\eqalign{S&={1 \over 2 \pi \alpha'} \int d^2z \Biggl[
\f{1}{2} \left[(\de+iu)Z_1\cdot (\bar{\de}-i\bar{u})\bar{Z}_1
+(\bar{\de}+i\bar{u})Z_1\cdot(\de-iu)\bar{Z}_1\right]\cr &\ \ +
\f12\left[(\de+iv)Z_2\cdot(\bar{\de}-i\bar{v})\bar{Z}_2
+(\bar{\de}+i\bar{v})Z_2\cdot(\de-iv)\bar{Z}_2\right]
+\f{1}{b}(\bar{u}v-\bar{v}u)
 \Biggr].}}
We still have not yet performed the path integral
over the zero-modes~\zeromode\ of $\varphi_1$, but if we now
combine those zero-modes together with the non-zero modes
of $\theta$ into a conventional field, then we can regard
$Z_1 = r_1 e^{i \theta}$ as a completely ordinary free field.

The final form~\pathfin\ is quadratic, so the one-loop functional
determinant is easily evaluated using the methods
of~\refs{\RussoCV,\RussoTJ, \RussoIK,\TakayanagiJJ}.  The result agrees
with~\superstring\ (when the
appropriate Green-Schwarz fermions are added to~\pathfin).
The integrals over the zero modes $u$, $v$
identified with the integrals over the auxiliary parameters
$\chi_1$, $\chi_2$ in~\superstring.

\subsec{Relation to $\IC^2/G$ Orbifolds}

We will now show that the result~\resone\ for the partition function
reduces to various orbifolds of $\IC^2$  with discrete torsion
when $\beta$ is rational.
In general, orbifolds with discrete torsion are defined by adding
extra phases $\ep$ in
the sum over twisted sectors
in the partition function \VafaWX.
The one-loop amplitude for a theory with abelian orbifold group $G$ can be
written as \eqn\onedis{Z=\f{1}{|G|}\sum_{g,h\in G}\ep(g,h)Z_{g,h}
.} Modular invariance requires that $\ep(g,h)$ should satisfy
the relations $\ep(g,hk)=\ep(g,h)\ep(g,k)$,
$\ep(g,h)=\ep(h,g)^{-1}$ and $\ep(g,g)=1$ \VafaWX.

Let us first consider the case $\beta = 4 /N$ such that $N$ and 4 are coprime.
Generalizing~\torusone, we can use the double periodicity~\periodicity\ to
divide both $\chi_i$ integrals,
\eqn\divideboth{
\eqalign{
Z &= \sum_{m,m',n,n' \in \IZ} \int_{T^2} {d^2 \chi_1 \over 2 \beta \tau_2}
\int_{T^2} {d^2 \chi_2 \over 2 \beta \tau_2}
e^{{\pi \over \tau_2 \beta} ( \chi_1 \bar{\chi}_2 - \chi_2 \bar{\chi}_1)}
e^{{8 \pi i \over \beta}(m' n - m n')}
\cr
&\qquad\qquad \times
e^{ { 2 \pi \over \tau_2 \beta}[ \chi_1(n' - n \bar{\tau}) -
\bar{\chi}_1 (n'-n \tau)]}
e^{{ 2 \pi \over \tau_2 \beta}[\chi_2 (m' - m \bar{\tau}) -
\bar{\chi}_2 (m'-m \tau)]}
F(\chi_1,\chi_2,\tau).
}}
The phase $e^{{8 \pi i \over \beta}(m'n - m n')}$ drops out
when $\beta = 4 /N$, and we can perform the summations
to obtain the delta-functions
\eqn\locz{ {1 \over 4} (\tau_2 \beta^2)^2 \sum_{p,p',q,q' \in \IZ}
\delta^2(\chi_1 - \ha \beta(p'-p \tau)) \delta^2(\chi_2 - \ha
\beta (q'-q\tau)). }
We see that the values of
$\chi_1$ and $\chi_2$ selected by the delta-functions~\locz\ are
those with $0 \le p,p',q,q' \le N - 1$.   In this case the
partition function becomes \eqn\orbione{ Z = {1 \over N^2}
\sum_{p,p',q,q' = 0}^{N-1} e^{{2 \pi i \over N} (p' q - p q')}
\left| { \vartheta_1(\ha (\nu_{p'p} + \nu_{q'q})|\tau)^2
\vartheta_1(\ha(\nu_{p'p} - \nu_{q'q})|\tau)^2 \over \eta(\tau)^6
\vartheta_1(\nu_{p'p}|\tau) \vartheta_1(\nu_{q'q}|\tau)}\right|^2,
} where we have defined $\nu_{m'm} = {2 \over N}(m' - m \tau)$.
We recognize \orbione\ as the usual orbifold partition function
for $\IC^2/\IZ_N \times \IZ_N$ defined by the two
$\IZ_N$ actions
\eqn\orbact{g_1:(X_1,X_2)\to (e^{4\pi i/N}X_1,X_2), \qquad
g_2:(X_1,X_2)\to (X_1,e^{4\pi i/N}X_2),}
multiplied by an
extra phase factor $e^{{2 \pi
i \over N} (p' q - p q')}$  which indicates the presence
of discrete torsion.

For $G=\IZ_N \times
\IZ_N$, the possible
phases which satisfy the consistency conditions listed below~\onedis\ 
are given by
\eqn\phaseids{\ep((m,l),(m',l'))=e^{\f{2\pi i}{N}(ml'-m'l)n},}
where $(m,l)$ denotes a boundary condition twisted by
$(g_1)^m(g_2)^l$. We have $N-1$ independent choices for the
discrete torsion phase
 $n=1,2,3,\ddd,N-1$ ($n=0$ corresponds to having no discrete
torsion).  Our result~\orbione\ for $\beta = 4/N$ corresponds
to the particular case $n=1$.

Similar results hold for more general rational values of $\beta$, though
we will only sketch the derivations, which are elementary.
There are four cases to consider, $\beta = \f{4k + l}{N}$ for
$l=0,1,2,3$.
For these general cases it
is convenient to start with the formula~\resone\ instead of~\divideboth.
For example, for $\beta = 4 k/N$ (i.e. $l=0$)
we can decompose the summation variable $p = N q + r$ (and  $p'$ similarly)
to write the sum over $p$ as
\eqn\trick{
\sum_{p \in \IZ} = \sum_{q \in \IZ} \sum_{r=0}^{N-1}.
}
The $\vartheta$-functions in~\resone\ are then independent of $q$ and $q'$.
Summing over these variables  gives delta-functions similar to~\locz\ which
localize the $\chi_2$ integral and give precisely the result~\orbione.

For $\beta = {4 k + 2 \over N}$ we must use the trick \trick\ with
$p = 2 N q + r$ (and let $r$ run from 0 to $2N-1$)
in order to decouple the sum over $q$, and for
$\beta={4k+1 \over N}$ or ${4k+3 \over N}$ we must use
$p = 4 N q + r$, with $r$ running from $0$ to $4N-1$.
At the end of the day, we find that
the four cases $l=0,1,2,3$ are respectively equivalent to the orbifolds
$\IC^2/\IZ_N \times \IZ_N$,
$\IC^2/\IZ_{4N}\times \IZ_{4N}$,
$\IC^2/\IZ_{2N}\times \IZ_{2N}$ and $\IC^2/\IZ_{4N}\times
\IZ_{4N}$ with discrete torsion, as summarized in~\results.
The discrete torsion in these four cases is given by~\phaseids\ with
$n$ defined by $kn = 1~{\rm mod}~N$,
$(4k+1)n=1~{\rm mod}~4N$,
$(2k+1)n=1~{\rm mod}~2N$,
or
$(4k+1)n=1~{\rm mod}~4N$, respectively.
They are non-supersymmetric orbifolds in type II string theory 
defined by  the actions of a $\beta\pi$ rotation in each complex plane.

At the values $\beta=\f{4k+1}{N},\f{4k+2}{N},\f{4k+3}{N}$ 
they can simultaneously be viewed as orbifolds with discrete torsion in
type 0 string theory.  This is because in these cases, the partition
function includes sectors with the (NS$-$,NS$-$) GSO projection.
 For example, when $\beta=2/N$ we find
\eqn\partonesscozn{Z=
{1 \over 4 N^2} \sum_{l,l',n,n'=0}^{2N-1} e^{\f{\pi
i}{N}(l' n-l n')}
\left |\f{\vartheta_1({1 \over 4}(\nu_{l'l}+\nu_{n'n})|\tau)^2
\vartheta_{1}({1 \over 4}(\nu_{l'l}-\nu_{n'n})|\tau)}{\eta(\tau)^{6}
\vartheta_1(\ha \nu_{l'l}|\tau) \vartheta_1(\ha \nu_{n'n}|\tau)}\right|^2.}
The terms in the sum with
$(l,n)=(N,0)$ or $(l,n)=(0,N)$ correspond to the (NS$-$,NS$-$) GSO
projection (this can be seen by using Jacobi's abstruse
identity).  These sectors
contain the usual bulk closed string tachyons of type 0 string theory.

\subsec{Bosonic String}

We can also consider the bosonic string in the sigma model~\sigmaone\ with
$D-4=22$ additional flat spacetime dimensions.
The only subtlety in calculating the partition function in the operator
formalism is the nontrivial zero-point energy.  We find
\eqn\aaa{
a = \tilde{a} = 1 + {1 \over 2} (\nu_1 - [\nu_1])^2 + {1 \over 2} (\nu_2
- [\nu_2])^2.
}
After a calculation similar to that in~\refs{\RussoCV,\RussoTJ,\HashimotoVE}
we find
the partition function
\eqn\aaa{
Z = \int { d^2 \chi_1\,d^2 \chi_2 \over (2 \beta \tau_2)^2}
e^{ {\pi \over \beta \tau_2} (\chi_1 \bar{\chi}_2 - \chi_2 \bar{\chi}_1)}
e^{-{\pi \over 2 \tau_2} (\chi_1 - \bar{\chi}_1)^2}
e^{-{\pi \over 2 \tau_2} (\chi_2 - \bar{\chi}_2)^2}
\left|{1 \over \eta(\tau)^{18} \vartheta_1(\chi_1|\tau)
\vartheta_2(\chi_2|\tau)} \right|^2.
}
Using manipulations similar to those above it is easy to show that
this partition function is invariant under
\eqn\aaa{
\beta \to \beta + 1,
}
as expected.
In the case of $\beta=k/N$, we always have the
the orbifold $\IC^2/\IZ_N\times \IZ_N$
with discrete torsion.

\newsec{D-branes}

In general, D-branes are important probes of
geometrical aspects of spacetime in string theory. Also they often
offer us various interesting models of Yang-Mills theory. In this
section we study D-branes in the background \backg. To define
D-branes, it is useful to analyze them in the free field representation
\actions, \pertw, where we can find an important class\foot{It is
possible that there exist more general D-brane boundary states
which cannot be obtained from this method. One of such examples
may be a D0-brane which is oscillating around the origin due to
the $r_{1,2}$ dependent dilaton.} of boundary states by imposing
Dirichlet or Neumann boundary conditions on these free fields
$X^1=r_1e^{i\phi_1}$ and $X^2=r_2 e^{i\phi_2}$.

We can consider the following nine possibilities for the boundary
conditions of $(r_1,\phi_1) $ and $ (r_2,\phi_2)$:
\eqn\poss{\eqalign{ & (a): (D,D) ,(D,D),\ \ \ (b): (N,N),(N,N),\ \
\ (c): (D,D),(N,N),  \cr & (d): (D,D),(N,D), \ \ \ (e):
(N,D),(N,D),\ \ \ (f): (N,D),(N,N), \ \ \  \cr & (c'):
(N,N),(D,D), \ \ \ (d'): (N,D),(D,D),\ \ \ (f'): (N,N), (N,D).}}
Notice that $(c'),(d'),(f')$ are essentially the same as
$(c),(d),(f)$ so we will not mention them.

\subsec{Toroidal D2-brane and Quantization of $b$}

First we discuss the $(a)$-type D-brane
in \poss, defined by
 imposing the Dirichlet boundary
condition in all directions $(X_1,\bar{X}_1,X_2,\bar{X}_2)$, i.e.
\eqn\bcondone{(\de-\bar{\de})\phi_{1,2}=(\de-\bar{\de})r_{1,2}=0.}
We can rewrite \bcondone\ in terms of the original sigma model coordinates
$\vp_{1}$ and $\vp_{2}$ using the formula (A.3) given in the
appendix A.  We find
\eqn\bcondtwo{\eqalign{(\de-\bar{\de})\phi_{1}
&=\f{br_2^2}{1+b^2r_1^2r_2^2}(\de+\bar{\de})\vp_2
+\f{1}{1+b^2r_1^2r_2^2}(\de-\bar{\de})\vp_1=0,\cr
(\de-\bar{\de})\phi_{2}
&=-\f{br_1^2}{1+b^2r_1^2r_2^2}(\de+\bar{\de})\vp_1
+\f{1}{1+b^2r_1^2r_2^2}(\de-\bar{\de})\vp_2=0.}} Comparing
\bcondtwo\ with the standard expression for the (mixed) Neumann
boundary condition
\eqn\bdrconmn{G_{\mu\nu}(\de+\bar{\de})X^{\nu}+(B_{\mu\nu}+2\pi
\al F_{\mu\nu})(\bar{\de}-\de)X^{\nu}=0,} we see that
this D-brane represents
a D2-brane whose world-volume is a torus $0\leq \vp_{1,2}\leq
2\pi$ at fixed values of $r_1$ and $r_2$. Moreover we find from~\bdrconmn\ 
that there is a non-zero gauge flux $F$ on this toroidal brane (these
computations are very similar to those in the Melvin
background \TakayanagiGU) \eqn\flux{2\pi\al
F_{\vp_1\vp_2}=\f{1}{b}.}
This D2-brane is stabilized by the presence of this flux, as we will
see shortly, even though it
wraps a topologically trivial cycle in spacetime.
Flux quantization requires
\eqn\fluxq{\f{1}{2\pi}\Tr\int d\vp_1 d\vp_2
\ F_{\vp_1\vp_2}=\f{k}{\beta}\in \IZ,} where $k$ is the number of
the D2-branes. Thus such a D-brane system is allowed when we have\foot{In this
analysis we do not see the modulo 4 distinction as in~\results.
This distinction comes
from the different spin structures of the superstring, but our
D0-brane analysis
is only sensitive to the bosonic degrees of freedom.  Indeed, in the bosonic
string, as we have seen in section 3.4, there is no distinction between
the four cases.}
\eqn\fcond{\beta=\f{k}{N}, } for an integer $N$.

When $\beta$ is irrational and $r_{1}$ and $r_2$ are
non-zero
this configuration is not consistent
with the quantization and thus does not exist. Still such a
D-brane can exist when $r_1=0$ or $r_2=0$. However, its
world-volume shrinks to zero size and it should be called a
D0-brane instead of a D2-brane, as is clear from~\bcondtwo.

This phenomenon can also be found from the effective theory
analysis. Consider $k$ D2-branes with magnetic flux $f=2\pi\al
F_{\vp_1\vp_2}$ wrapped on a torus $0\leq \vp_1,\vp_2\leq 2\pi$
for fixed values of $(r_1,r_2)$
 in the background \backg.   We can find its energy from the DBI action
 \eqn\dbidt{\eqalign{M_{N,f}&=\f{e^{-\phi}}{4\pi^2(\al)^{3/2}}
 \Tr\int d\vp_1d\vp_2 \s{\det(G+B+2\pi\al F)}
 \cr &=k\f{e^{-\phi_0}}{\al^{3/2}}\s{r_1^2r_2^2(fb-1)^2+f^2}.}}
 For generic values of $f$, the energy
 of the D2-branes is only stabilized at $r_1 r_2=0$, where
 it should be regarded as a D0-brane. In the rational case $f=1/b=\al
 N/k$, however, there exist stable D2-brane configurations for any non-zero
values of
 $r_1$ and $r_2$. These values precisely
 agree with \flux\ and should be quantized according to \fcond.
 Then the system can be regarded as
 a bound state of $N$ D0-branes and $k$ D2-branes. Due to the
 magnetic flux, the open string theory becomes non-commutative.
Following the standard formula~\SeibergVS
\eqn\noncomm{
\theta^{ij} = 2 \pi \alpha' (G + B + 2 \pi \alpha' F)_{\rm asymmetric}^{-1~ij}
}
we can see that the world-volume of the D2-brane becomes a
 non-commutative torus with rational 
 non-commutativity parameter\foot{We define the non-commutative torus with
 the non-commutativity $2\pi \beta$ by the algebra $UV=VUe^{2\pi
 i\beta}$.} $2\pi\beta=2\pi k/N$, i.e. a fuzzy torus.
 On the other hand, the backgrounds with
irrational values of $\beta$ can be regarded as an $N\to \infty$
limit, in which case the mass of D2-brane becomes infinite. These
correspond to a non-commutative torus with irrational
non-commutativity.

In the rational case, we have seen the model \sigmaone\ becomes equivalent to
the orbifold $\IC^2/\IZ_N\times \IZ_N$ with discrete torsion.
In this model, following general arguments
\refs{\DouglasXA,\DouglasHQ} (see also appendix C), we have two kinds of
D0-branes.
One is a regular bulk D0-brane, which has moduli to freely move around
the four-dimensional space.
The other is called a fractional D0-brane, which is stuck on
the fixed lines $r_1r_2=0$ and whose mass is $1/N$ times smaller
than that of a bulk D0-brane.
 It is now clear that the  toroidal
D2-brane and the localized D0-brane that we find in the background \backg\ are
equivalent to the bulk D0-brane and fractional D0-brane,
respectively.
Another check of this fact is that the mass \dbidt\ of the D2-brane in
the rational case becomes
\eqn\massff{M_{N,f}=N\f{e^{-\phi_0}}{\s{\al}},}
which is
precisely
the same as that of $N$ D0-branes on the fixed lines.

We would also like to point out that the properties of a fractional
D0-brane in orbifolds with discrete torsion are actually not the
same as those in ordinary orbifolds without discrete torsion. In
the latter case, the brane cannot move away from the fixed points
(or lines) at all even if we push it by some external force. On
the other hand, in the former case, it can be pushed off the fixed points
by adding extra force.
This shows that the fixed points (or lines)
are not rigid in the presence of discrete torsion.  This fact
is also clear from the geometry~\backg, in which the geometry around
$r_1 r_2 = 0$ is smoothed out.

\subsec{Classification of D-branes}

So far we have only analyzed case $(a)$ in~\poss.
We can analyze the other D-branes $(b)$---$(f)$ in the same way
as~\bcondtwo\ using the transformations given in appendix A.
 We omit the details of this analysis and
summarize the results. The D-branes for cases $(d),(e)$ and $(f)$ seem to exist
only when $\beta$ is rational as will be discussed employing the
boundary state analysis in the next subsection.

$(a)$: D0-brane located on the fixed lines $r_1r_2=0$. In the
     rational case, we can also

     obtain a D2-brane wrapped
     on the torus with gauge flux $2\pi\al F_{\vp_1\vp_2}=\f{1}{b}$.

$(b)$: D4-brane filling the whole spacetime (no flux).

$(c)$: D2-brane filling the plane $X_2,\bar{X}_2$ (no flux).

$(d)$: D3-brane (= $S^1\times \IC$; $r_1=$fixed) with flux
$2\pi\al F_{\vp_1\vp_2}=\f{1}{b}$. This brane

exists only in the
   rational case.

$(e)$: D4-brane filling the whole spacetime with  flux $2\pi\al
F_{\vp_1\vp_2}=\f{1}{b}$. This brane

exists only in the
   rational case.

$(f)$: D3-brane ($\vp_1=$ fixed, i.e. $\IR^3$) with no flux.
  This brane exists only in the

   rational case.

\subsec{Boundary States}

To describe D-branes including $\al$ corrections, it is useful to
construct boundary states. At  rational values of $\beta$, as
we have seen in section 3, the theory is equivalent to the
orbifold with discrete torsion $\IC^2/\IZ_N \times \IZ_N$ (see
also appendix C).
In this case we can compare the D-brane
spectrum with the expected result.

First, let us construct the boundary states $|B\lb$ for $(a),(b)$
and $(c)$.
In these cases, we can show that the identity
\eqn\idenfP{(J_{i}+\ti{J}_{i})|B\lb=J_{i0}|B\lb, \qquad i=1,2,}  
holds, where $J_{i0}$
denotes the zero mode part of the angular momentum
 $J_{i0}= i\s{\f{2}{\al}}(x_{i0}\bar{\ap}_{i0}-\bar{x}_{i0}\ap_{i0})$.
A similar identity holds in the Melvin background~\TakayanagiAJ,
and in this subsection we follow
 conventions of that paper.
In other words, these boundary conditions (i.e. (purely) Neumann
or Dirichlet in each $X^i$ direction) ensure that the oscillator
parts of the boundary states do not have any $J_{i}$ charges.

In the $(b)$ D-brane case, the zero mode parts $J_{i0}$ are also
zero due to the Neumann boundary condition. Thus its boundary
state is the same as the familiar one in the flat space. At
rational $\beta$, it is equivalent to a (bulk) D4-brane in the
orbifold theory with discrete torsion.

The $(a)$ D-brane boundary states have non-zero values of $J_{i0}$
in general. It looks impossible to write down boundary states when
$J_{i0}$ is an arbitrary integer due to the self-interacting mode
shifts \pertw. However, if we assume that either $r_1$ or $r_2$ is
zero, we have $\hat{J}_{10}=0$ or $\hat{J}_{20}=0$. Then we can write
down the boundary state (we consider the $r_1=0$ case and suppress
fermions for simplicity) \eqn\boundarys{|B\lb = \sum_{J_{20}\in\IZ
}\exp\left[\sum_{n=0}^{\infty}\f{1}{n+\nu_1}
\ap^{1}_{-n-\nu_1}\bar{\ti{\ap}}^1_{-n-\nu_1}+\f{1}{n}\ap^{2}_{-n}
\bar{\ti{\ap}}^2_{-n}+h.c.\right] |J_{10}=0,J_{20}
\lb_{\rm zeromode},} up to a normalization factor which can
be determined by  open-closed duality. This D-brane has the
moduli to move in the $X^2$ direction. The other one located at
$r_2=0$ can be found in the same way. When $\beta$ is rational, we
can classify the values of $J_{20}$ by the integers mod $N$
($m=0,1,2,\ddd, N-1$)
\eqn\jtre{ |J_{20}=m \lb_{\rm zeromode}=\f{1}{N}
\sum_{l=0}^{N-1}e^{\f{2\pi i}{N}(\hat{J}_2-m)l} |X_2
\lb_{\rm zeromode}.} Then the boundary state \boundarys\
 agrees with the $\IC^2$ analogue of the boundary state
 for a fractional D0-brane
computed in \CrapsXW\ for the orbifold $\IC^3/\IZ_N\times \IZ_N$ .
In the rational case we can also define a D0-brane which can
freely move in any direction by projecting the angular momenta to
$\hat{J}_{1,2}\in N\IZ$. This can be done by acting with the operators
$\Omega_i=\sum_{l=0}^{N-1}e^{2\pi i \hat{J}_{i} l/N}$ on the
boundary states \boundarys. Its boundary state only includes the
untwisted sector and is equivalent to a bulk (or regular) D0-brane
in the orbifold theory. This construction manifestly shows that
the bulk D0-brane is made of $N$ fractional D0-branes at $N$
different positions.

The boundary state for a $(c)$ D-brane can be constructed almost
in the same way as  the $(a)$ case.
This corresponds to the (bulk) D2-brane in the orbifold theory.

The analysis of boundary state for cases $(d),(e),(f)$ is a bit
different from the previous ones. We  no longer have the relation
\idenfP\ due to the mixed ND boundary condition. However, we can
still project the whole boundary states by the operators
$\Omega_i$ when $\beta$ is rational. Notice that  now they act
also on the massive oscillators. Putting this operator  in the ND
part, we can straightforwardly find the boundary states for
$(d),(e)$ and $(f)$. They are a respectively
the bulk D1, D2 and D3-branes in the
orbifold theory.

\newsec{Three Parameter Model}

In this section we consider the generalization
of~\backg\ to a three parameter deformation of $\IC^3$.
Similar to the AdS case studied in~\LuninJY,
we can make a sequence of transformations,
$(TsT)_{b_1} (TsT)_{b_2} (TsT)_{b_3} $ of flat space to obtain the background
\FrolovDJ
\eqn\aaa{\eqalign{
&ds^2=\sum_{i=1}^3 (dr_i^2+g r_i^2 d\phi_i)+g r_1^2 r_2^2 r_3^2
(\sum_{i=1}^3 b_i d\phi_i)^2,
\cr
&
B=g (b_3 r_1^2 r_2^2 d\phi_1 \wedge d\phi_2+b_1 r_2^2 r_3^2 d\phi_2
\wedge d\phi_3+b_2 r_3^2 r_1^2 d\phi_3 \wedge d\phi_1),
\cr
&
e^{2 \phi}=e^{2 \phi_0} g,~~~~g^{-1}=1+b_3^2 r_1^2 r_2^2+b_1^2 r_2^2
r_3^2+b_2^2 r_3^2 r_1^2.
}}
Below we will use different angular variables related to $\phi_i$ by
\eqn\variables{
\psi = {1 \over 3} (\phi_1 + \phi_2 + \phi_3),\qquad
\varphi_1 = {1 \over 3} (\phi_2 + \phi_3 - 2 \phi_1), \qquad
\varphi_2 = {1 \over 3} (\phi_1 + \phi_3 - 2 \phi_2).
}
The background preserves supersymmetry when $|b_1| = |b_2| = |b_3|$,
as we will see shortly.

\subsec{Partition Function}

As shown in~\RussoYU, the sigma model for closed strings in this
background can be solved with the same kind of transformations
we reviewed in section 2.
In the canonical quantization,
one has
\eqn\Ldefsthree{\eqalign{
L_0 &=  N - a -\sum_{k=1}^3 (\nu_k -
[\nu_k]) J_k , \cr \tilde{L}{}_0 &= 
\tilde{N} - \tilde{a} + \sum_{k=1}^3 (\nu_k - [\nu_k])
\tilde{J}_k, }} where \eqn\nudefs{ \nu_i = \epsilon_{ijk} (J_j +
\tilde{J}_j) \beta_k\ \ \ \ \ (\beta_i\equiv b_i\al). }
We can immediately write down the analogue of \oneloop\ and
\steptwo, \eqn\oneloopp{ Z_{T^2}=iV_{D-6} \int \f{d\tau
d\bar{\tau}}{4\tau_2} {1 \over (4\pi^2 \alpha' \tau_2)^{\ha(D-6)}}
Z(\tau,\bar{\tau},\beta), } and
\eqn\aaa{\eqalign{ Z &= \int
d^2j_1\,d^2j_2\,d^2j_3\,d^2\chi_1\, d^2\chi_2\,d^2 \chi_3 \cr
&e^{-2 \pi i (\chi_1 j_1 + \bar{\chi}_1
\bar{\jmath}_1 + \chi_2 j_2 + \bar{\chi}_2 \bar{\jmath}_2 + \chi_3
j_3 + \bar{\chi}_3 \bar{\jmath}_3)} e^{2 \pi i \tau[ - \nu_1  j_1
- \nu_2 j_2 - \nu_3  j_3 ]} e^{-2 \pi i \bar{\tau}[ \nu_1
\bar{\jmath}_1 + \nu_2  \bar{\jmath}_2 + \nu_3 \bar{\jmath}_3 ]}
\cr &\qquad\times \left|{ \vartheta_1(\ha \chi_{+++}|\tau)
\vartheta_1(\ha \chi_{++-}|\tau) \vartheta_1(\ha \chi_{+-+}|\tau)
\vartheta_1(\ha \chi_{+--}|\tau) \over \eta(\tau)^3
\vartheta_1(\chi_1|\tau) \vartheta_1(\chi_2|\tau)
\vartheta_1(\chi_3|\tau)} \right|^2, }} where \eqn\aaa{
\chi_{s_1 s_2 s_3} = s_1 \chi_1 + s_2 \chi_2 + s_3 \chi_3. }
Plugging in \nudefs\ and performing the $j$ integrals gives
\eqn\aaase{\eqalign{ Z &= \int {d^2\chi_1\, d^2\chi_2\,d^2 \chi_3
\over (2 \tau_2)^2}
\delta^2(\beta_1 \chi_1 +
\beta_2 \chi_2 + \beta_3 \chi_3, \beta_1 \bar{\chi}_1 + \beta_2
\bar{\chi}_2 + \beta_3 \bar{\chi}_3)\cr
& \times e^{{\pi \over \beta_3
\tau_2}(\chi_1 \bar{\chi}_2 - \chi_2 \bar{\chi}_1)} 
\left|{ \vartheta_1(\ha \chi_{+++}|\tau)
\vartheta_1(\ha \chi_{++-}|\tau) \vartheta_1(\ha \chi_{+-+}|\tau)
\vartheta_1(\ha \chi_{+--}|\tau) \over \eta(\tau)^3
\vartheta_1(\chi_1|\tau) \vartheta_1(\chi_2|\tau)
\vartheta_1(\chi_3|\tau)} \right|^2. }}
Recalling that $\vartheta_1(0|\tau) = 0$, it is 
manifest that~\aaase\ vanishes when
$|b_1|=|b_2|=|b_3|$, indicating the presence of unbroken
spacetime supersymmetry.  This is similar to the mechanism
of~\refs{\TakayanagiJJ,\RussoNA}.

One can also derive this result from a path integral calculation
using the world-sheet action \eqn\pathfinthr{\eqalign{S&=
{1 \over 2\pi\al}\int d^2z
\Biggl[\f{1}{b_3}(\bar{u}v-\bar{v}u)+\f{1}{2}
\left[(\de+iu)Z_1\cdot (\bar{\de}-i\bar{u})\bar{Z}_1
+(\bar{\de}+i\bar{u})Z_1\cdot(\de-iu)\bar{Z}_1\right]\cr &\ \ +
\f12\left[(\de+iv)Z_2\cdot(\bar{\de}-i\bar{v})\bar{Z}_2
+(\bar{\de}+i\bar{v})Z_2\cdot(\de-iv)\bar{Z}_2\right]\cr &\ \
+\f12\Bigl[(\de-i(b_1u+b_2v)/b_3)Z_3
\cdot(\bar{\de}+i(b_1\bar{u}+b_2\bar{v})/b_3)\bar{Z}_3 \cr & \ \
+(\bar{\de}-i(b_1\bar{u}+b_2\bar{v})/b_3)Z_3
\cdot(\de+i(b_1u+b_2v)/b_3)\bar{Z}_3\Bigr]\Biggr].}} It is
straightforward to check that \pathfinthr\ becomes the same
sigma model defined from the three parameter model \aaa\ after
integrating out $u$ and $v$, and that its one-loop functional
determinant agrees with~\aaase.

We can study the properties of the partition function \aaase\ as
in section 3. Consider the supersymmetric case
$\beta_1=\beta_2=\beta_3(\equiv \beta)$. Then we can show the
(formal) periodicity $Z(\beta+1,\tau,\bar{\tau})=Z(\beta,\tau,\bar{\tau})$
in the same way as in \betashift. In the rational case $\beta=k/N$
we find that the partition function is equivalent to that of the
orbifold $\IC ^3/\IZ_N \times \IZ_N$ with discrete torsion. The
orbifold actions are defined by \eqn\orbasusy{\eqalign{&
g_1:(X_1,X_2,X_3)\to (e^{2\pi ik/N}X_1,X_2,e^{-2\pi ik/N}X_3),\cr &
g_2:(X_1,X_2,X_3)\to (X_1,e^{2\pi ik/N}X_2,e^{-2\pi ik/N}X_3).}}

\subsec{D-branes}

The analysis of D-branes in this background can be done as in
section 4.
We continue to consider only the supersymmetric case
$\beta_1=\beta_2=\beta_3(=\beta)$ and only discuss the
supersymmetric D0-brane in
 the free field description (the analogue of case (a) from section 4).
In this example, the Dirichlet condition
$(\de-\bar{\de})\vp'_1=(\de-\bar{\de})\vp'_2=(\de-\bar{\de})\psi=0$
can be rewritten using the formula (A.6) into
\eqn\condthree{\eqalign{&(r_1^2+r_3^2)(\de+\bar{\de})\vp_1+
(r_3^2-r_1^2)(\de+\bar{\de})\psi+r_3^3(\de+\bar{\de})\vp_2
-b^{-1}(\de-\bar{\de})\vp_2=0, \cr &
(r_2^2+r_3^2)(\de+\bar{\de})\vp_2+
(r_3^2-r_2^2)(\de+\bar{\de})\psi+r_3^3(\de+\bar{\de})\vp_1
+b^{-1}(\de-\bar{\de})\vp_1=0, \cr & (\de-\bar{\de})\psi=0.}}

Comparing \condthree\ with the standard formula as before we find
a D2-brane wrapped on the  torus $(\vp_1,\vp_2)$ at any fixed
values of $\psi,r_{1},r_2$ and $r_3$ with gauge flux $2\pi \al
F_{\vp_1,\vp_2}=1/b$. Again the rational values $\beta(=\al
b)=k/N$ are required for flux quantization. This flux
together with the background B-field makes its world-volume a
non-commutative torus with  non-commutativity $2\pi\beta$ (see
also appendix C). This fact nicely agrees with the
$\beta$ deformation of ${\cal N}=4$ Yang-Mills theory \refs{\LuninJY,
\BerensteinUX,\BerensteinHY}
 realized on D5-branes wrapped on the
torus.

\newsec{Time-Dependent Background via $\beta$ Deformation}

We can also study $\beta$ deformed $\IR^{1,3}$ by
replacing one of the two complex free fields
$(X^1,\bar{X}^1)$ in \sigmaone\ with a Lorentzian one
$(X^{+},X^{-})$ with metric  $ds^2=-dX^+ dX^-$. We
leave $(X^2,\bar{X}^2)$ unchanged. Define polar coordinates by
$X^+=t e^{\theta'},\ \ X^-=t e^{-\theta'}$ and $X^{2}=r e^{i\phi}$
$(-\infty< t <\infty, 0\leq r\leq \infty)$. These coordinates
cover one half of the four dimensional Minkowski spacetime
$\IR^{1,3}$.

Now we perform T-duality in the $\theta'$ direction, shift
by $\phi=\vp_2+b\ti{\theta}$, and  T-dualize  in the
$\ti{\theta}$ direction again. Then we get the following
time-dependent background\foot{This can also be obtained from the
analytic continuation $t\to it,\ \theta\to i\theta$ and $b\to
ib$ of \backg.} \eqn\backg{\eqalign{
ds^2&=-dt^2+dr^2+\f{t^2}{1+b^2 t^2 r^2}d\theta^2+
\f{r^2}{1+b^2 t^2 r^2}d\vp^2,  \cr
e^{2(\phi-\phi_0)}&=\f{1}{1+b^2 t^2 r^2}, \cr B_{\theta
\vp}&=-\f{b t^2 r^2}{1+b^2 t^2 r^2}.}} The angular coordinate
$\vp$ is periodic $\vp\sim \vp+2\pi$ as usual. The boost
coordinate $\theta$ can be compact or non-compact. Below we assume
that $\theta$ is compact with $\theta\sim \theta+2\pi$ and also that
$\beta=\alpha' b$ is rational (=$k/N$). In the trivial case $\beta=0$, the
spacetime is the product of the Milne orbifold
$[\IR^{1,1}/\IZ]_{\Delta=2\pi}$ (i.e. the orbifold by the boost
$\Delta=2\pi$) and the flat space $\IC$. For recent development of
orbifold theoretic approaches to time-dependent backgrounds see
\CornalbaKD\ and references therein.

When $\beta \neq 0$ the spacetime shows an intriguing time-evolution.
At $t=-\infty$, the spacetime is given by the orbifold
$[\IR^{1,1}/\IZ]_{\Delta=2\pi/N}\times [\IC/\IZ_N]$. As time
passes, a bubble of the new geometry which looks like
$[\IR^{1,1}/\IZ]_{\Delta=2\pi}\times \IC$ is created at the origin
$r=0$ and its radius expands as $r\sim \f{1}{b|t|}$. At $t=0$ it
covers the whole region completely and thus the spacetime becomes
$[\IR^{1,1}/\IZ]_{\Delta=2\pi}\times \IC$. After that, the
time-reversal process occurs and the spacetime goes back to the
original orbifold.

This time evolution can be regarded as a time-dependent process
of resolving an orbifold space and is very similar to the
scenario\foot{For time-dependent supergravity solutions for the
decay of $\IC/\IZ_N$ refer to \refs{\GregoryYB,\HeadrickYU}.} in
localized closed string tachyon condensation on the
non-supersymmetric orbifold $\IC/\IZ_N$ \AdamsSV. It is expected
that this time-dependent model can be exactly solvable, as in the
space-like case. We leave the detailed study of this model for future work.

\centerline{\bf Acknowledgments}

We are grateful to A.~Adams for pointing out that the orbifolds under
consideration here have closed string tachyons.
This research was supported in part by the National Science Foundation under
Grant No.~PHY99-07949. AV acknowledges support from the U. S. Department
of Energy under grant number DE-FG02-90ER40542.

\vskip 0.5in

\appendix{A}{T-duality Relations}

Here we summarize the on-shell relations between the various fields 
employed in section 2, using the rule~\tduality.
{}From~\sigmaone\  to~\sigmatwo\ we find
\eqn\transfot{\eqalign{\de\tvp_1&=-\f{br_1^2r_2^2}{1+b^2r_1^2r_2^2}\de\vp_2
-\f{r_1^2}{1+b^2r_1^2r_2^2}\de\vp_1, \cr
\bar{\de}\tvp_1&=-\f{br_1^2r_2^2}{1+b^2r_1^2r_2^2}\bar{\de}\vp_2
+\f{r_1^2}{1+b^2r_1^2r_2^2}\bar{\de}\vp_1.}}
Then from~\sigmatwo\  to~\sigmathree\ we find
\eqn\transftt{
\eqalign{
\de\phi_1
&
=-\f{1}{r_1^2}\de\tvp_1,
\cr
\qquad
\bar{\de}\phi_1
&
=\f{1}{r_1^2}\bar{\de}\tvp_1.}
}
Combining these results, we see that in
going from~\sigmaone\ to~\sigmathree\ 
we have the transformation
\eqn\transfotr{\eqalign{\de\phi_1&=\f{br_2^2}{1+b^2r_1^2r_2^2}\de\vp_2
+\f{1}{1+b^2r_1^2r_2^2}\de\vp_1, \cr
\bar{\de}\phi_1&=-\f{br_2^2}{1+b^2r_1^2r_2^2}\bar{\de}\vp_2
+\f{1}{1+b^2r_1^2r_2^2}\bar{\de}\vp_1.}}

In the same way we find the following three relations for $\tilde{\varphi}_2$,
which was defined in section 2 in a manner similar to $\tvp_1$:
\eqn\transfot{\eqalign{\de\tvp_2&=\f{br_1^2r_2^2}{1+b^2r_1^2r_2^2}\de\vp_1
-\f{r_2^2}{1+b^2r_1^2r_2^2}\de\vp_2, \cr
\bar{\de}\tvp_2&=\f{br_1^2r_2^2}{1+b^2r_1^2r_2^2}\bar{\de}\vp_1
+\f{r_2^2}{1+b^2r_1^2r_2^2}\bar{\de}\vp_2,\cr
\de\phi_2&=-\f{1}{r_2^2}\de\tvp_2,\cr
\bar{\de}\phi_2&=\f{1}{r_2^2}\bar{\de}\tvp_2,\cr
\de\phi_2&=-\f{br_1^2}{1+b^2r_1^2r_2^2}\de\vp_1
+\f{1}{1+b^2r_1^2r_2^2}\de\vp_2, \cr
\bar{\de}\phi_2&=\f{br_1^2}{1+b^2r_1^2r_2^2}\bar{\de}\vp_1
+\f{1}{1+b^2r_1^2r_2^2}\bar{\de}\vp_2.}}
{}From the above relations we can show
that
\eqn\relddd{\de\phi_1=\de(\vp_1-b\tvp_2),\ \ \
\bar{\de}\phi_1=\bar{\de}(\vp_1-b\tvp_2).}
Thus $\phi_1=\vp_1-b\tvp_2$. Also we have
$\phi_2=\vp_2+b\tvp_1$ by definition.

For the supersymmetric ($b_i = b$ for $i=1,2,3$) model considered
in section 5 we find
\eqn\relationtr{\eqalign{ \de \vp'_1&=\f{1+b r_3^2}{1+b^2
g_0}\de\vp_1 +\f{b(r_3^2+r_2^2)}{1+b^2 g_0}\de \vp_2
+\f{b^2[r_1^2r_2^2+r_1^2r_3^2-2r_2^2 r_3^2]+b(r_3^2-r_2^2)}{1+b^2
g_0}\de\psi,\cr \bar{\de} \vp'_1&=\f{1-b r_3^2}{1+b^2
g_0}\bar{\de}\vp_1 -\f{b(r_3^2+r_2^2)} {1+b^2 g_0}\bar{\de} \vp_2
+\f{b^2[r_1^2r_2^2+r_1^2r_3^2-2r_2^2 r_3^2]-b(r_3^2-r_2^2)}{1+b^2
g_0}\bar{\de}\psi, \cr \de \vp'_2&=\f{1-b r_3^2}{1+b^2
g_0}\de\vp_2 -\f{b(r_3^2+r_1^2)}{1+b^2 g_0}\de \vp_1
+\f{b^2[r_1^2r_2^2+r_1^2r_3^2-2r_2^2 r_3^2]-b(r_3^2-r_1^2)}{1+b^2
g_0}\de\psi,\cr \bar{\de} \vp'_2&=\f{1+b r_3^2}{1+b^2
g_0}\bar{\de}\vp_2 +\f{b(r_3^2+r_1^2)} {1+b^2 g_0}\bar{\de} \vp_1
+\f{b^2[r_1^2r_2^2+r_1^2r_3^2-2r_2^2 r_3^2]+b(r_3^2-r_1^2)}{1+b^2
g_0}\bar{\de}\psi .}}

\appendix{B}{Partition Functions}

We record here some
formulas which are useful for computing the partition functions
appearing in section 3.
We use the notation
\eqn\aaa{
q = e^{2 \pi i \tau}, \qquad y = e^{2 \pi i a}, \qquad
z = e^{2 \pi i b}.
}

Consider first the bosonic oscillators.  In $D$ spacetime
dimensions there are a total of $D-2$ physical bosons, with four
of them describing the sigma model~\sigmaone\ and the remaining
$D-6$ free.  The partition function in the bosonic
Hilbert space weighted
by the angular momenta $J_1$ and $J_2$ is
\eqn\Zboson{
\eqalign{
Z_{\rm B} &\equiv \Tr_{\rm B}[q^N y^{J_1} z^{J_2}]\cr
&=
{y^{-\ha} \over 1 - 1/y} {z^{-\ha} \over 1 - 1/z}
\prod_{n=1}^\infty
{1 \over (1 - q^n)^{D-6}}
{1 \over 1 - q^n y}
{1 \over 1 - q^n/y}
{1 \over 1 - q^n z}
{1 \over 1 - q^n/z}
\cr
&= -
{q^{(D - 2)/24} \eta(\tau)^{8 - D}
\over \vartheta_{1}(a|\tau) \vartheta_{1}(b|\tau)}.
}}
The corresponding partition function for fermions in the Ramond
sector is
\eqn\aaa{\eqalign{
Z_{\rm R} &\equiv
\Tr_{\rm R}[q^N y^{J_1} z^{J_2}]
\cr
&= 4 (y^{\ha} + y^{-\ha})(z^\ha + z^{-\ha}) \prod_{n=1}^\infty
(1 + q^n)^4 (1 + q^n y) (1 + q^n/y)
(1 + q^n z) (1 + q^n/z)
\cr
&= q^{-1/3} {\vartheta_{2}(0|\tau)^2 \vartheta_{2}(a|\tau)
\vartheta_{2}(b|\tau) \over \eta(\tau)^4}.
}}
The projection onto states of definite fermion number $e^{\pi i F} = \pm 1$
gives, for either sign, just a factor of $1/2$,
\eqn\aaa{
Z_{{\rm R} \pm} = {1 \over 2} Z_{\rm R}.
}
For the NS sector, we have
\eqn\aaa{\eqalign{
Z_{\rm NS} &\equiv q^{-\ha} \Tr_{\rm NS}[q^{N}
y^{J_1} z^{J_2}]
\cr
&= q^{-\ha} \prod_{n=1}^\infty
(1 + q^{n - \ha})^4 (1 + q^{n - \ha} y) (1 + q^{n - \ha}/y)
(1 + q^{n - \ha} z) (1 + q^{n - \ha}/z)
\cr
&= q^{-1/3} {\vartheta_{3}(0|\tau)^2
\vartheta_{3}(a|\tau) \vartheta_{3}(b|\tau)
\over
\eta(\tau)^4
}.
}}
Note that we have already included here the ground state energy
$a=-1/2$, which accounts for the prefactor  $q^{-1/2}$ in the first line.
The projection onto definite fermion number $e^{\pi i F} = \pm 1$ now
gives
\eqn\aaa{
Z_{{\rm NS}\pm } = {q^{-1/3} \over 2 \eta(\tau)^4}
\left[
\vartheta_{3}(0|\tau)^2
\vartheta_{3}(a|\tau) \vartheta_{3}(b|\tau)
\mp \vartheta_{4}(0|\tau)^2
\vartheta_{4}(a|\tau) \vartheta_{4}(b|\tau)
\right].
}
The type II GSO projection
is given by
\eqn\aaa{
\eqalign{
Z_{{\rm NS+}} - Z_{{\rm R}\pm}
&= {q^{-1/3} \over 2 \eta(\tau)^4}
\sum_{i=2}^4 (-1)^{i+1} \vartheta_i(0|\tau)^2 \vartheta_i(a|\tau)
\vartheta_i(b|\tau)\cr
&=
{q^{-1/3} \over \eta(\tau)^4}
\vartheta_1(\half(a+b)|\tau)^2
\vartheta_1(\half(a-b)|\tau)^2,
}}
using Jacobi's abstruse identity.
The $\pm$ subscript on $Z_{\rm R}$ indicates type IIA/IIB, which are
equivalent in this context.

\appendix{C}{D-branes on Orbifolds with Discrete Torsion}
 
In orbifolds with discrete
torsion, we can analyze D-branes by following the general
procedures found in \refs{\DouglasXA,\DouglasHQ}. Let the action of $g\in G$ on
the Chan-Paton factor be denoted by $\gamma(g)$. The projection
in the open string theory is given by
\eqn\projopen{\phi=\gamma(g)^{-1}\cdot ( g\cdot\phi)\cdot
\gamma(g),} as in the usual orbifold without discrete torsion
\DouglasSW. In the presence of discrete torsion, $\gamma(g)$
provides a projective representation of $G$. In particular, the
consistency of open and closed string theory  gives the relation
\refs{\DouglasXA,\DouglasHQ,\GomisEJ}
\eqn\reqoec{\gamma(g)\gamma(h)
=\ep(g,h)\gamma(h)\gamma(g).}

Let us now recall the result we found for the case $\beta = 4/N$,
which was given by \phaseids\ with $n=1$.
In this case \reqoec\ becomes the fuzzy torus algebra generated by
$U=\gamma(1,0)$ and $V=\gamma(0,1)$, which satisfy $UV=VUe^{2\pi
i/N}$. This has the standard $N\times N$ matrix representation.
After imposing the orbifold projection \projopen\ , we find that
the two transverse scalars $\Phi_1$ and $\Phi_2$ should be
proportional to $V^{2}$ and $U^{-2}$, respectively.  D-branes
defined in this
 way are called fractional
 D-branes \refs{\DiaconescuBR,\DouglasXA,\DouglasHQ,\DiaconescuDT,
\GomisEJ}.  Note that an $N\times N$ Chan-Paton matrix is employed to describe
 a fractional D-brane. For its boundary state refer to \CrapsXW.

 Let us consider  fractional D0-branes. Since 
 the twisted sectors for $g_1$ (or $g_2$) are included, there are
no zero-modes for $X_1$ (or $X_2$)
 and they
 cannot freely move in those directions.
 The non-abelian gauge theory on  $N$ such D0-branes can be
 realized by multiplying another
 Chan-Paton degree of freedom with this $N\times N$ matrix.
 When we make the latter matrix
 implicit, we will have a non-commutative deformation. The commutator
 $[\Phi_1,\Phi_2]=\Phi_1\Phi_2-\Phi_2\Phi_1$ is $\beta$ deformed as
 \eqn\betadf{[\Phi_1,\Phi_2]_{\beta}=e^{\pi i\beta}\Phi_1\Phi_2-
 e^{-\pi i \beta}\Phi_2\Phi_1,}
 where $\beta=4/N$ as expected. It is natural to believe this
 deformation continues to be true when $\beta$ is irrational,
in which case the boundary state analysis shows that there also exist
fractional D0-branes.
When the Chan-Paton matrix has size $N$ and $\Phi_1$ and $\Phi_2$
are respectively proportional to $V^2$ and $U^2$,
we see that they have vanishing $\beta$ deformed commutator.
 This represents
a regular (or bulk) D0-brane which can move freely in any directions.

 Even though we analyzed D-branes in the particular case
 $\IC^2/\IZ_N \times \IZ_N$,
 we can obtain almost the same results for the supersymmetric orbifold
 $\IC^3/\IZ_N \times \IZ_N$ \BerensteinHY. This is equivalent to the
 three parameter model
we study in section 5
 at $\beta_i=1/N$.

\listrefs

\end